\documentclass[ALICE,manyauthors]{cernphprep}

\usepackage[comma,square,numbers,sort&compress]{natbib}

\usepackage[T1]{fontenc} % if needed
\usepackage[utf8]{inputenc}
\usepackage[english]{babel}
\usepackage{graphicx,subfigure}
\usepackage{amssymb}
\usepackage{hyperref}
\usepackage{lineno}
\usepackage[amssymb]{SIunits}
\usepackage{booktabs}
\usepackage{multirow}
\usepackage{xcolor}
\usepackage{overpic}
\usepackage{units}
\usepackage{placeins}
\newcommand{\pp}{\ensuremath{\mbox{p\kern-0.05em p}}}
\newcommand{\ppbar}{\ensuremath{\mathrm{p\kern-0.05em \bar{p}}}}
\newcommand{\pPb}{\ensuremath{\mbox{p--Pb}}}

\newcommand{\sqrtS}{\ensuremath{\sqrt{s}}}
\newcommand{\sqrtSnn}{\ensuremath{\sqrt{s_{\mathrm{NN}}}}}
\newcommand{\sqrtSE}[2][TeV]{\ensuremath{\sqrtS = #2~\mathrm{#1}}}
\newcommand{\sqrtSnnE}[2][TeV]{\ensuremath{\sqrtSnn = #2~\mathrm{#1}}}
\newcommand{\GeVc}{\ensuremath{\mathrm{GeV}\kern-0.05em/\kern-0.02em \textit{c}}}
\newcommand{\gev}{\ensuremath{\mathrm{GeV}\kern-0.05em}}
\newcommand{\ptjet}{\ensuremath{p_\mathrm{T,\,jet}}}
\newcommand{\pttrack}{\ensuremath{p_\mathrm{T,\,track}}}
\newcommand{\pttrigger}{\ensuremath{p_\mathrm{T,\,trigger}}}
\newcommand{\jt}{j_\mathrm{T}}
\newcommand{\jtch}{j_\mathrm{T,\,ch}}
\def\pt#1{\ensuremath{p_{\rm T#1}}}
\def\vjt#1{\ensuremath{\vec{j}_{\rm T#1}}}
\def\kt#1{\ensuremath{k_{\rm T#1}}}

\def\rms#1{\ensuremath{\sqrt{\left<#1^2\right>}}}

\begin{document}

%%%%%%%%%%%%%%%  Title page %%%%%%%%%%%%%%%%%%%%%%%%
\begin{titlepage}
\PHyear{2020}
\PHnumber{220}      % required, will be obtained from PH
%\PHdate{6 September}  % required, will be obtained from PH
\PHdate{10 November}  % required, will be obtained from PH
%

%%% Put your own title + short title here:
\title{Jet fragmentation transverse momentum distributions \\in pp and $\pPb$ collisions at \sqrtS, $\sqrtSnnE{5.02}$}
\ShortTitle{Jet fragmentation transverse momentum}   % appears on right page headers

%%% Do not change the next lines
\Collaboration{ALICE Collaboration\thanks{See Appendix~\ref{app:collab} for the list of collaboration members}}
\ShortAuthor{ALICE Collaboration} % appears on left page headers, do not change
%\linenumbers
\begin{abstract}
Jet fragmentation transverse momentum ($\jt$) distributions are measured in proton-proton (pp) and proton-lead (\pPb) collisions at $\sqrtSnnE{5.02}$ with the ALICE experiment at the LHC. Jets are reconstructed with the ALICE tracking detectors and electromagnetic calorimeter using the anti-$\kt{}$ algorithm with resolution parameter $R=0.4$ in the pseudorapidity range $|\eta|<0.25$. The $\jt$ values are calculated for charged particles inside a fixed cone with a radius $R = 0.4$ around the reconstructed jet axis. 
The measured $\jt$ distributions are compared with a variety of parton-shower models. Herwig and \textsc{Pythia} 8 based models describe the data well for the higher $\jt$ region, while they underestimate the lower $\jt$ region. 
The $\jt$ distributions are further characterised by fitting them with a function composed of an inverse gamma function for higher $\jt$ values (called the ``wide component''), related to the perturbative component of the fragmentation process, and with a Gaussian for lower $\jt$ values (called the ``narrow component''), predominantly connected to the hadronisation process. The width of the Gaussian has only a weak dependence on jet transverse momentum, while that of the inverse gamma function increases with increasing jet transverse momentum. For the narrow component, the measured trends are successfully described by all models except for Herwig. For the wide component, Herwig and PYTHIA 8 based models slightly underestimate the data for the higher jet transverse momentum region. These measurements set constraints on models of jet fragmentation and hadronisation.
\end{abstract}
\end{titlepage}
\setcounter{page}{2}

%%%%% Body of the article

% !TEX root = main.tex

%% main text
\section{Introduction}
\label{sec:introduction}
Jets are groups of collimated particles mainly resulting from fragmentation of hard scattered partons produced in high-energy particle collisions.
Jet production in quantum chromodynamics (QCD)~\cite{gross1973ultraviolet, politzer1973reliable,gross1973asymptotically, gross1974asymptotically, georgi1974electroproduction} can be thought as a two-stage process~\cite{eventGenerators}. After being produced in the hard scattering, partons reduce their virtuality by emitting gluons~\cite{basicsofpqcd}. Since the momentum transfer scale ($Q^{2}$) is large during the showering, perturbative QCD calculations can be applied. When $Q^{2}$ becomes of the order of $\Lambda_{\mathrm{QCD}}$, partons hadronise into final-state particles through processes that cannot be calculated perturbatively~\cite{introPythia81,herwigManual,herwig7releaseNote,nmlla,Kang:2017glf,Azimov:1984np,Gutierrez-Reyes:2018qez}. 
Instead, the implementation of specific hadronisation models in Monte Carlo event generators such as PYTHIA~\cite{introPythia81} and Herwig~\cite{herwig7releaseNote} can be used.

In this work the fragmentation of partons is studied using the jet fragmentation transverse momentum, $\jt$. The $\jt$ is defined as the perpendicular component of the momentum of the constituent particle with respect to reconstructed jet momentum, $\vec{p}_{\mathrm{jet}}$. The length of the $\vjt{}$ vector is
  \begin{equation}
    \jt = \frac{|\vec{p}_{\mathrm{jet}} \times \vec{p}_{\mathrm{track}}|}{|\vec{p}_{\mathrm{jet}}|} \quad,
  \label{eq:jtdefinition}
  \end{equation}
where $\vec{p}_{\mathrm{track}}$ is the momentum of the constituent particles. It is one of many jet shape observables to study the properties of fragmenting particles with respect to the initial hard momentum during the fragmentation process. The $\jt$ provides a measurement of the transverse momentum spread of the jet fragments.

Previously, $\jt$ has been studied using two-particle correlations where $\jt$ is calculated for particles with respect to the highest transverse momentum particle in each event instead of reconstructed jet.
The study using the correlation method was done by the CCOR collaboration at ISR in $\pp$ collisions at centre-of-mass energies $\sqrtS = 31,\;45$ and $63~\mathrm{GeV}$~\cite{firstjtmeasurement} and by the PHENIX collaboration at RHIC in $\pp$ collisions at $\sqrtSE[GeV]{200}$~\cite{PHENIXjets} and d--Au collisions at a center-of-mass energy per nucleon pair $\sqrtSnnE[GeV]{200}$~\cite{phenixJtPAu}. The results showed no clear dependence on the transverse momentum ($p_\mathrm{T}$) of the trigger particle. 
Jet measurements to study $\jt$ were done by the CDF collaboration in $\ppbar$ collisions at $\sqrtSE{1.96}$~\cite{cdfpaper} at Tevatron, by the ATLAS collaboration in $\pp$ at $\sqrtSE[TeV]{7}$~\cite{Aad:2011sc} 
and by the LHCb collaboration in $\pp$ collisions at $\sqrtSE[TeV]{8}$~\cite{Aaij:2019ctd} at the LHC. The results show a dependence of the width of $\jt$ distributions with respect to the $p_\mathrm{T}$ of jets at the LHC energies. 

Jets are used as an important probe for the study of the deconfined phase of strongly interacting matter, the quark--gluon plasma (QGP) that is formed in high-energy collisions of heavy nuclei.
There exists plenty of experimental evidence of jet energy loss, such as the suppression of inclusive hadron spectra at high transverse momentum~\cite{Adcox:2001jp,Adams:2003im,Arsene:2003yk,Khachatryan:2016odn,Acharya:2018qsh}, the modification of back-to-back hadron-hadron~\cite{Adare:2007vu,Aamodt:2011vg} and direct photon-hadron correlations~\cite{Adare:2012qi}, hadron--jet correlations ~\cite{Adam:2015doa,Adamczyk:2017yhe}, the modification of reconstructed jet spectra~\cite{Adam:2015ewa,Acharya:2019jyg} and jet substructure~\cite{Sirunyan:2018qec,Chatrchyan:2014ava,Acharya:2018uvf,Acharya:2019djg}, as compared to the expectations from elementary proton-proton collisions.

Jet quenching in heavy-ion collisions evolves multi-scale steps from hard to soft processes~\cite{Kurkela:2014tla,Tachibana:2018yae}. Hard scales dominate in the elementary hard scattering. The hard scattering is followed by the subsequent branching process down to non-perturbative scales. Soft scales, of the order of the temperature of the medium, characterise interactions of soft partons produced in the shower with the QGP. Soft scales also govern hadronisation, which is expected to take place in vacuum for sufficiently energetic probes, even though some changes can persist from modifications of colour flow~\cite{Aurenche:2011rd,Beraudo:2011bh,Beraudo:2012bq}. Understanding the contributions from the different processes to the jet shower evolution in medium and their scale dependence is crucial to constrain the dynamics of jet energy loss in the expanding medium~\cite{CasalderreySolana:2012ef}, and fundamental medium properties like the temperature-dependent transport coefficient~\cite{DEramo:2012uzl,Ayala:2016pvm}. 
Besides heavy-ion collisions one should study also smaller systems such as p--Pb in order to get an important baseline.
Cold nuclear matter effects~\cite{Baier:1996sk,McLerran:1993ni,Eskola:2009uj} in p--Pb collisions need to be considered to interpret the measurements in heavy-ion collisions.   

The results for $\jt$ distributions obtained using two-particle correlations were recently reported by the ALICE Collaboration~\cite{Acharya:2018edi}  in $\pp$  and $\pPb$ collisions. In this paper, 
jet reconstruction provides a better estimate of the initial parton momentum than the leading hadron in two-particle correlations. 
Additionally, contrary to the correlation studies, the $\jt$ distribution is not smeared by hadrons decaying from a short living resonance.

The $\jt$ distributions are studied by reconstructing jets with the ALICE tracking detectors and electromagnetic calorimeter using the anti-$\kt{}$ algorithm~\cite{antikt} with resolution parameter $R=0.4$ in the pseudorapidity range $|\eta|<0.25$ in pp collisions at $\sqrtS$ = 5.02 TeV and $\pPb$ minimum bias collisions at $\sqrtSnnE{5.02}$. It is worth noting that there is a shift in the centre-of-mass rapidity of $\Delta y = 0.465$ in the direction of the proton beam because of the asymmetric collision system.
The $\jt$ distribution is further analysed by fitting and separating it into two distinct components that are assigned to the parton shower and the hadronisation process. The attempt to separate the two components is presented for the first time using jets in various jet transverse momentum (\ptjet) ranges.
We also compare the results with those obtained from PYTHIA (PYTHIA 8.3) and Herwig (Herwig 7.2) simulations.

\section{Experimental setup and data samples}
\label{sec:experimentaldetails}
The data presented here were recorded by the ALICE detector in 2017 for pp collisions at $\sqrtS = 5.02\,\mathrm{TeV}$ with $7.6 \times 10^8$  minimum-bias events ($\mathcal{L}_{\mathrm{int}} = 15.7\,\mathrm{nb}^{-1}$) and in 2013 for $\pPb$ collisions at $\sqrtSnnE{5.02}$ with $1.3 \times 10^{8}$ events ($\mathcal{L}_{\mathrm{int}} = 620\,\mathrm{nb}^{-1}$). Detailed information about the ALICE detector during LHC Run 1 and Run 2 can be found in Refs.~\cite{aliceDetector,alicePerformance}.

The V0 detector~\cite{forwarddetectorsTdr} provides the information for event triggering. The V0 detector consists of two scintillator hodoscopes that are located on each side of the interaction point along the beam direction. It covers the pseudorapidity region $-3.7 < \eta < -1.7$ (V0C) and $2.8 < \eta < 5.1$ (V0A). To select the minimum-bias trigger signals are required in both the V0A and V0C . This condition is used to reduce the contamination of data from beam-gas events using the timing difference of the signals between the V0A and V0C detectors~\cite{alicePerformance}.

The analysis is performed with events that have a primary vertex within $|z_\mathrm{vtx}|<10~\mathrm{cm}$ of the nominal interaction point at $z_\mathrm{vtx}=0$ along the beam direction. Charged particles are used for reconstruction of the primary vertex and jets. The charged particles are reconstructed with the Inner Tracking System (ITS)~\cite{aliceITS} and the Time Projection Chamber (TPC)~\cite{aliceTPC}. These detectors are located inside a large solenoidal magnet that provides a homogeneous magnetic field of \unit[0.5]{T}. Tracks within a pseudorapidity range $|\eta| < 0.9$ over the full azimuth are accepted. 
The ITS is made up of the Silicon Pixel Detector (SPD) in the innermost layers, the Silicon Drift Detector (SDD) in the middle layers and the Silicon Strip Detector (SSD) in the outermost layers, each consisting of two layers.
The tracks are selected following the hybrid approach~\cite{hybridExplanation} which ensures a uniform distribution of tracks as a function of azimuthal angle ($\varphi$). 
The hybrid approach combines two different classes of tracks. The first class consists of tracks that have at least one hit in the SPD. The tracks from the second class do not have any SPD associated hit and mainly rely on the position information of the primary vertex when reconstructing the tracks. Combining the information from the ITS and TPC provides a $\pt{}$ resolution ranging from $1$ to $10\,\%$ for charged particles from $0.15$ and $\unit[100]{\GeVc}$. For tracks without the ITS information, the momentum resolution is comparable to that of ITS+TPC tracks below transverse momentum $\pt{}=\unit[10]{\GeVc}$, but for higher momenta the resolution reaches $20\,\%$ at $\pt{}=\unit[50]{\GeVc}$~\cite{alicePerformance,aliceBackgroundFluctuation}.

The EMCal covers an area with a range of $|\eta|<0.7$  in pseudorapidity and 107 degrees in azimuth and is made up of 12288 towers in total. Each tower consists of 76 alternating layers of \unit[1.44]{mm} lead and 77 layers of \unit[1.76]{mm} scintillator material.
The EMCal is also used to provide a high-energy photon trigger for a high-\ptjet\ data sample that is complementary to the minimum bias trigger for a low \ptjet\ data sample. The EMCal can be used to trigger on single shower deposits or energy deposits integrated over a larger area. The latter is used for the high-energy photon trigger. The EMCal trigger definition for $\pPb$ collisions in 2013 requires an energy deposit in a group of the towers of either \unit[10]{~\gev}  for the low threshold trigger or \unit[20]{~\gev} for the high threshold trigger. A sample of ~$3\times10^6$ events (  $\mathcal{L}_{\mathrm{int}}~=~5\,\mathrm{nb}^{-1}$) with the EMCal trigger provides increased statistics for $\ptjet>60{~\GeVc}$ where the trigger bias disappears in the analysis~\cite{Abelev:2014ffa}.  
The energy of the electromagnetic shower clusters is reconstructed in the EMCal by searching for a tower with an energy deposit greater than a defined seed energy and merging all towers that share the energy cluster.
To avoid double counting, when a cluster is matched with charged particles measured by the ITS and TPC, the sum of the transverse momentum of all the matched tracks are subtracted from the cluster energy.

\section{Analysis method}
\label{sec:methods}
For each collision event, jets are reconstructed with the anti-$\kt{}$ algorithm~\cite{antikt} and resolution parameter $R=0.4$ using FastJet~\cite{fastjet}. The $p_\mathrm{T}$-recombination scheme is used when reconstructing jets. Jets are selected in $\left| \eta \right| < 0.25 $ to satisfy the fiducial acceptance of the EMCal. 
The jet energy resolution $\mathrm{JER} = \sigma(p_\mathrm{T, jet}^\mathrm{reco})/p_\mathrm{T, jet}^\mathrm{true}$ is calculated as 20\% (18\%) at $p_\mathrm{T, jet}^\mathrm{true} = 20$ GeV/$c$ and 21\% (19\%) at 100 GeV/$c$ in pp (p--Pb) collisions. The jet angular resolution is estimated as 29\% (28\%) and 2\% (2\%) at $\ptjet = 20$ GeV/c  20\% (19\%) and 1.2\% (1.2\%) at $\ptjet = 100$ GeV/c in pp (p--Pb) collisions for pseudorapidity and azimuthal angle, respectively.  In the jet reconstruction both charged particles with $\pt{}>\unit[0.15]{\GeVc}$ and EMCal clusters with $\pt{}>\unit[0.3]{\GeVc}$ are considered. All charged particles within a fixed cone with a resolution parameter $R$ are taken as jet constituents, instead of using the list of jet constituents provided by the jet algorithm~\cite{Aad:2011sc,Chatrchyan:2012mec}. Results are presented in terms of the jet transverse momentum \ptjet.

The resulting $\jt$ distributions are corrected for detector effects using the unfolding method in Ref.~\cite{Adye:2011gm}. The response matrix used for the unfolding is obtained from events generated by PYTHIA 8 Monash 2013 (PYTHIA 8.2)~\cite{Skands:2014pea} for the correction of the data sample in pp collisions and PYTHIA 6 Perugia 2011 (PYTHIA 6.4)~\cite{Skands:2010ak} for the correction of the one in p--Pb collisions. The events are transported through the ALICE experimental set up described with GEANT 3~\cite{Agostinelli:2002hh,Asai:2015xno}. This response matrix $( j_\mathrm{T}^\mathrm{rec}, p_\mathrm{T, jet}^\mathrm{rec}, j_\mathrm{T}^\mathrm{true}, p_\mathrm{T, jet}^\mathrm{true} )$ has $2\times2$ dimensions to correct the detector inefficiency for jet transverse momentum ($\ptjet$) and $\jt$ simultaneously, where $j_\mathrm{T}^\mathrm{true}$ and $p_\mathrm{T, jet}^\mathrm{true}$ are obtained from particle level jets by PYTHIA 6 and 8 and $j_\mathrm{T}^\mathrm{rec}$ and $p_\mathrm{T, jet}^\mathrm{rec}$ are the corresponding measured values in ALICE, respectively. As a primary method the unfolding is performed with an iterative (Bayesian) algorithm as implemented in the RooUnfold package~\cite{Adye:2011gm}. The unfolding procedure is tested by dividing the generated data sample into two halves. The first half is used to fill the response matrix. The second half is used to test the closure of the unfolding method. For $40<\ptjet<\unit[150]{\GeVc}$, the generated $\ptjet$ distribution is recovered. For $\jt>\unit[0.1]{\GeVc}$, the $\jt$ distribution is also recovered.    

The effect of the underlying event background is estimated by looking at a cone perpendicular to the observed jet axis ($\frac{\pi}{2}$ rotation in $\varphi$, for details see Refs~\cite{ALICE:2014dla,Acharya:2018eat}). The background $\jt$ is calculated for any track that is found within this cone and the rotated jet axis is used as reference for $\jt$. The background obtained in this manner is subtracted from the unfolded inclusive $\jt$ distribution, which gives the resulting signal distribution as shown in Eq.~\ref{eq:inclbg}. The probability of events with jets inside the perpendicular cone are estimated as 1--2\% of the total number of jets. Jets reconstructed with charged particles only (charged jet) for $R=0.4$ and $p_\mathrm{T,\,jet}^\mathrm{ch}>5$ GeV/$c$ are used to check other jets inside the perpendicular since charged jets can cover the full azimuthal angle contrary to the case of jets in the EMCal acceptance. To make sure there is no jet contribution in the background, those events are not used for background estimation. Because of this reason, $N_{\textrm{perpendicular jets}}$ is less than $N_{\textrm{jets}}$ by about 1--2\% in Eq.~\ref{eq:inclbg}. 
\begin{equation}
 \frac{1}{N_{\mathrm{jets}}}\frac{\mathrm{d}N}{\jtch \mathrm{d} \jtch}\bigg\vert_\mathrm{signal} = \frac{1}{N_{\mathrm{jets}}}\frac{\mathrm{d}N}{\jtch \mathrm{d} \jtch}\bigg\vert_\mathrm{inclusive} - \frac{1}{N_{\textrm{perpendicular jets}}}\frac{\mathrm{d}N}{\jtch \mathrm{d}  \jtch}\bigg\vert_\mathrm{background} 
\label{eq:inclbg}
\end{equation}

The resulting signal distribution is fitted with the two-component function shown in Eq.~\ref{eq:fit}. 
A Gaussian distribution centered at $\jt~=~\unit[0]~{\GeVc}$ is used for lower $\jt$ and an inverse gamma function is used for $\jt$ above $\unit[1]{\GeVc}$, where $B_1$ to $B_5$ are parameters~\cite{Acharya:2018edi}. 
\begin{equation}
\frac{1}{N_{\mathrm{jets}}}\frac{\mathrm{d}N}{\jtch \mathrm{d} \jtch} = \frac{B_2}{B_1\sqrt{2\pi}}e^{-\frac{\jt^2}{2B_1^2}}+\frac{B_3B_5^{B_4}}{\Gamma\left(B_4\right)}\frac{e^{-\frac{B_5}{\jt}}}{\jt^{B_4+1}}
\label{eq:fit}
\end{equation}
To achieve stable results the fitting is performed in two steps. First, lower and higher parts of the $\jt$ distribution are fitted with a Gaussian and inverse gamma function, respectively. After getting the results from the individual fits, they are combined into a single function with initial values from the individual results and then an additional fit is performed. 
After getting the fit function, $\sqrt{\left<\jt^2\right>}$ (RMS) and yield values are extracted separately from each component. The narrow component RMS from the Gaussian part is determined as
\begin{equation}
\sqrt{\left<\jt^2\right>}=\sqrt{2}B_1
\label{eq:rmsnarrow}
\end{equation}
and the wide component RMS value from the inverse gamma function is calculated as
\begin{equation}
\sqrt{\left<\jt^2\right>}=\frac{B_5}{\sqrt{\left(B_4-2\right)\left(B_4-3\right)}}\quad,
\label{eq:rmswide}
\end{equation}
where it is required that $B_4 > 3$.

\section{Systematic uncertainties}
\label{sec:systematicerrors}
The systematic uncertainties in this analysis come from the background estimation, the unfolding procedure and the uncertainties related to track and cluster selection. The effect originating from uncertainty in the tracking efficiency is estimated with a PYTHIA simulation by removing 4\% of tracks randomly from each event corresponding to a mismatching probability of tracks between the ITS and TPC. The resulting variations in the RMS values are less than 4\% and 5\% for the wide and narrow components, respectively. The uncertainty related to the EMCal energy scale was estimated by scaling cluster energies up and down by 2\% in the PYTHIA particle level generation in order to reflect a non-linearity correction of the EMCal energy scale ranging from about 7\% at $\unit[0.5]{\GeVc}$ to a negligible value above $\unit[3]{\GeVc}$. Similarly, the jet momentum was scaled by $\pm2\%$ when determining $\ptjet$ to check how the cluster energy affects $\jt$ distributions. The variation of both RMS components is seen to be less than $2\%$. 

The systematic uncertainty on the background estimation was studied using the ``random background'' method as an alternative to that of the perpendicular-cone. This method assigns new random $\eta$ and $\varphi$ of the existing tracks in the event using a uniform distribution without changing their $p_\mathrm{T}$ values. A random jet cone is also from uniform $\eta$ and $\varphi$ distributions covering $|\eta|<0.25$ and $0<\varphi<2\pi$ and tracks near the jet axis are not used. The resulting uncertainty is below 5\% for the wide component RMS and below 9\% for the narrow component RMS in p--Pb collisions. 
%In pp collisions, the background estimation with the random background method is overestimated as the underlying events are expected to be smaller than the p--Pb data. As a result, the signal $\jt$ distribution is underestimated. 
To study the effect of background fluctuations in p--Pb collisions, a study based on embedding particles generated with PYTHIA in real events was performed. The embedded particles are simulated by following the multiplicity density information~\cite{ALICE:2012xs} and $\pt{}$ distribution~\cite{Abelev:2014dsa} of charged particles in p--Pb collisions in ALICE. The effect in RMS is negligible for both RMS components.  

The systematic uncertainty introduced by the unfolding procedure was determined by repeating the unfolding using the Singular-Value Decomposition (SVD) method as an alternative~\cite{Hocker:1995kb}. Given that the SVD method does not allow for multi-dimensional unfolding, the unfolding is performed separately for different \ptjet\ intervals. In a PYTHIA closure test, the true distribution for $\jt>0.1\,\mathrm{GeV}/c$ was in general found to be between the unfolded distributions from the iterative and SVD methods within 2\%. The difference between the methods when unfolding data is used as an estimate of the unfolding uncertainty. The iterative unfolding algorithm permits the change of the number of iterations as a regularisation parameter. The stability of the results was verified by using one iteration above and below instead of the default value, where the default value is chosen by checking that unfolded $\jt$ distributions converge. Also, the regularisation parameter $k$ is varied by one unit above and below with respect to the default solution of the SVD method that is determined by following the guideline~\cite{Hocker:1995kb}. The iterative algorithm requires a prior estimate of the shape of the distribution.
As a default prior, generated PYTHIA distribution is used. To estimate the effect of the prior, the unfolded $\jt$ distribution is used as a prior instead. The effect of the unfolding for different ranges of $\ptjet$ is tested by varying the first value of $\ptjet$ from 5 to \unit[15]{GeV/$c$}. These effects are found negligible compared to that for the two different unfolding methods. The resulting uncertainty by the unfolding procedure is below 8\% for both wide and narrow component RMS in p--Pb collisions. In pp collisions it is 9\% and 12\% for the wide and narrow components, respectively.

The model dependence of the unfolding procedure was explored by weighting the response matrix with PYTHIA. The jet yield in the response matrix is varied by $\pm30\%$ for the angularity $g > 0.1$. The angularity is defined as $g = \Sigma_i \left( p_{\mathrm{T},i} \times r_i \right) / p_\mathrm{T,jet}$, where $p_{\mathrm{T},i}$ is the $\pt{}$ of the $i^\mathrm{th}$ constituent of the jet and $r_i = \sqrt{\Delta \eta^2_i + \Delta \varphi^2_i}$ is the distance of the $i^\mathrm{th}$ constituent from the jet axis~\cite{Larkoski:2014pca,Acharya:2019jyg}. The effect is found to be below 2\% for the wide component and negligible for the narrow component.

The different sources of systematic uncertainty are considered as uncorrelated and the values are summed in quadrature. The summary table in Table~\ref{tab:systematics} shows an overview of systematic uncertainties for $40<\ptjet<60$ GeV/$c$ in pp and p--Pb collisions.

\begin{table}[htb]
\centering
\caption{Summary of systematic uncertainties for $40<\ptjet<60$ GeV/$c$ in pp and p--Pb collisions.}
\label{tab:systematics}
\begin{tabular}{ c | c | c |  c | c | c | c }
   &  \multicolumn{2}{c|}{$\jt$ distribution at $\jt$ = 0.2--0.8--\unit[2]{GeV/$c$}} &  \multicolumn{2}{c|}{Wide RMS} &   \multicolumn{2}{c}{Narrow RMS} \\
  \hline			
  source   &   pp  &  p--Pb  &  pp  & p--Pb  & pp   & p--Pb \\
  \hline
  Background & 2--2--5\% & neg.--2--5\% & 1.1\%  & 5\% & 2.9\%  & 9\%   \\
  Unfolding & 10--neg.--20\% & 10--neg.--12\% & 9\% & 8\%& 12\% & 8\%   \\
  Tracking & 2--2--2\% & 2--1--neg.\% & 0.4\% & 4 \% & 0.2\% &  5\%    \\ 
  EMCal & 2--2--5\% & 2--2--2\% & 1.8\% &  1\% & 0.2\% & 1\%  \\
  Model dependence & neg.--2--5\% & neg.--neg.--10\% & 0.5\% & 2\% & neg. & neg.   \\
  \hline
  Total & 11--4--22\%&  10--3--16\% & 9\% &  10 \% & 12\% &  13\% \\
  \hline
  \end{tabular}
\end{table}

\section{Results}
\label{sec:results}

%%%%%%%%%%%%% RESULTS %%%%%%%%%%%%%%%%%%%%%%%%%
\begin{figure}[!htb]
  \begin{center}
  \includegraphics[width=0.8\textwidth]{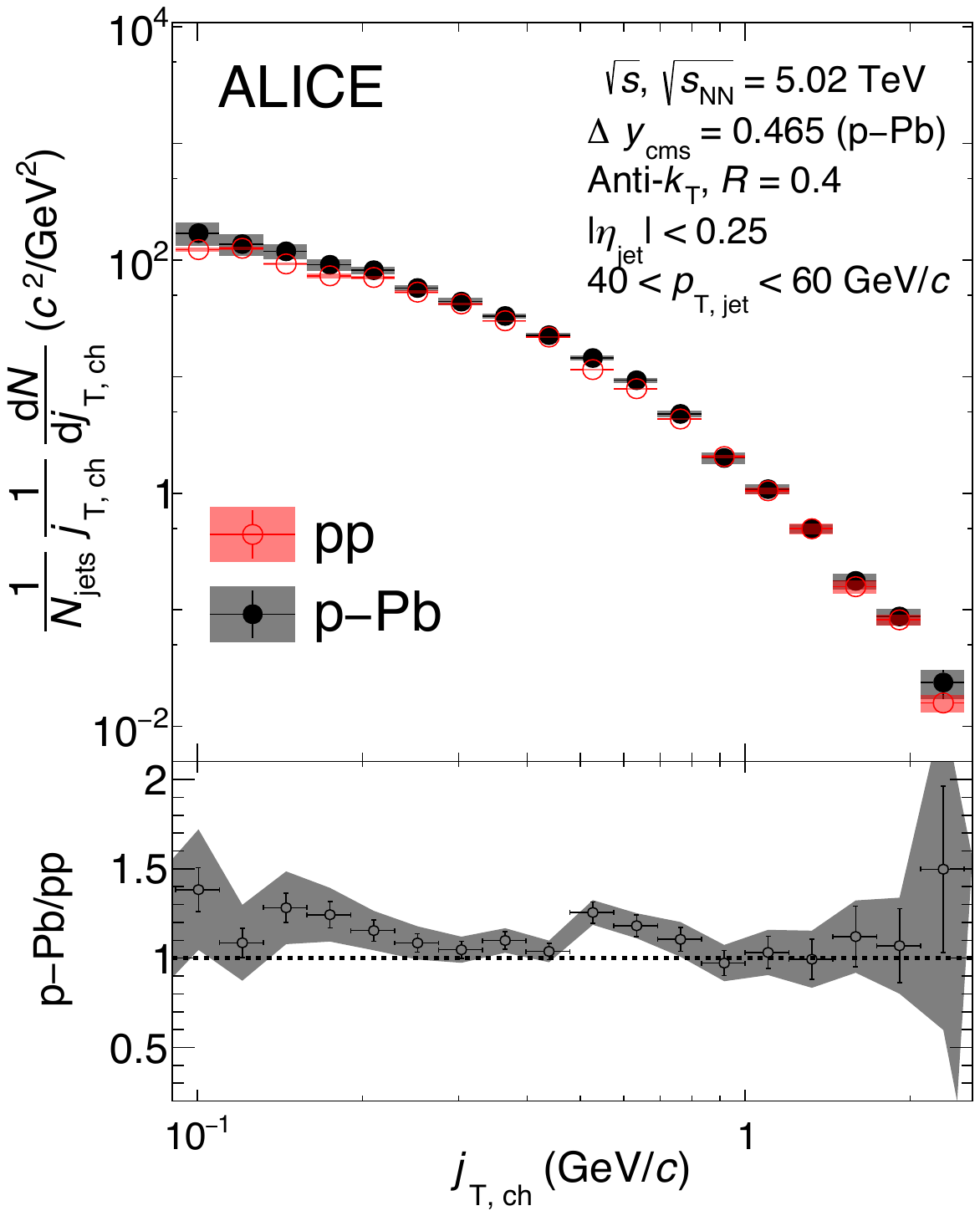}
  \caption{Comparison of the $\jt$ distributions in pp and $\pPb$ collisions at $\sqrtS$, $\sqrtSnn$ = 5.02 TeV in $40<\ptjet<60$ GeV/$c$. The centre-of-mass rapidity in p--Pb collisions is shifted by $\Delta y = 0.465$ in the direction of the proton beam.} 
  \label{fig:jtppboverpp}
  %Fig2_DrawFinalFits.py in https://github.com/TWSman/JtAnalysis
  \end{center}
\end{figure}

The $\jt$ distribution in pp collisions at $\sqrtS = 5.02\,\mathrm{TeV}$ is compared with that in p--Pb collisions at $\sqrt{s_\mathrm{NN}} = 5.02\,\mathrm{TeV}$ in Fig.~\ref{fig:jtppboverpp} for jet transverse momentum in  \unit[$40<\ptjet<60$]{GeV/$c$}. The ratio of the $\jt$ distributions represents the consistence of the result in pp and p--Pb collisions and implies no clear cold nuclear matter effects in p--Pb collisions.  For the interval in $100<\ptjet<150$ GeV/$c$, the comparison is not provided because of the lack of enough statistics in minimum-bias pp collisions and the absence of the data sample with the EMCal trigger in the corresponding pp data taking period. 

\begin{figure}[!htb]
  \begin{center}
  \includegraphics[width=0.8\textwidth]{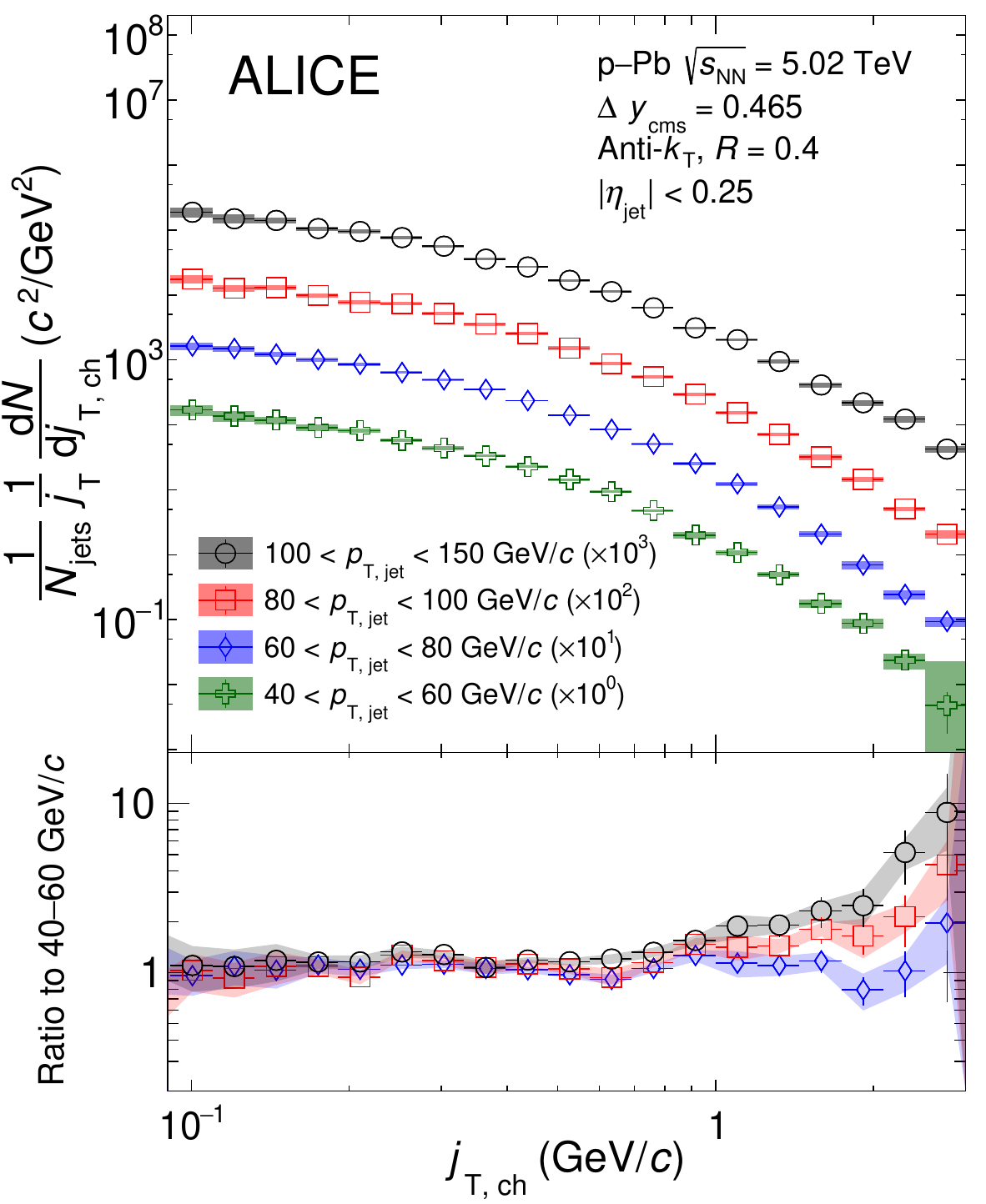}
  \caption{The $\jt$ distributions of charged particles in $R = 0.4$ anti-$k_\mathrm{T}$ jets as measured in $\pPb$ collisions at $\sqrtSnnE{5.02}$ for different ranges of $\ptjet$. The centre-of-mass rapidity is shifted by $\Delta y = 0.465$ in the direction of the proton beam. The bottom panel shows ratios of the $\jt$ distributions with respect to that in \unit[$40<\ptjet<60$]{GeV/$c$}. }
  \label{fig:jt_systematics}
  \end{center}
  \end{figure}

Figure~\ref{fig:jt_systematics} shows the distributions of $\jt$ for charged particles in different $\ptjet$ intervals after applying the unfolding correction and background subtraction in $\pPb$ collisions at \sqrtSnnE[TeV]{5.02}.
The yield at low $\jt$ stays constant with increasing $\ptjet$. At high $\jt$ the yield increases and the distributions become wider with increasing $\ptjet$ as indicated by the ratios of the $\jt$ distributions shown in the bottom panel. Notably, this is due to kinematical limits. At midrapidity, within a fixed cone the maximum $\jt$ depends on the track momentum by the relation of $j_\mathrm{T,\,max} \approx R \times \pttrack$, resulting in an increase of the possible $\jt$ as $\ptjet$ increases. Though jets with larger momenta are more collimated, the net effect is an increase of $\langle\jt\rangle$ as $\ptjet$ increases. These measurements are consistent with the findings by the ATLAS~\cite{Aad:2011sc} and LHCb collaborations~\cite{Aaij:2019ctd}.

\begin{figure}[t]
  \begin{center}
  \includegraphics[width=0.8\textwidth]{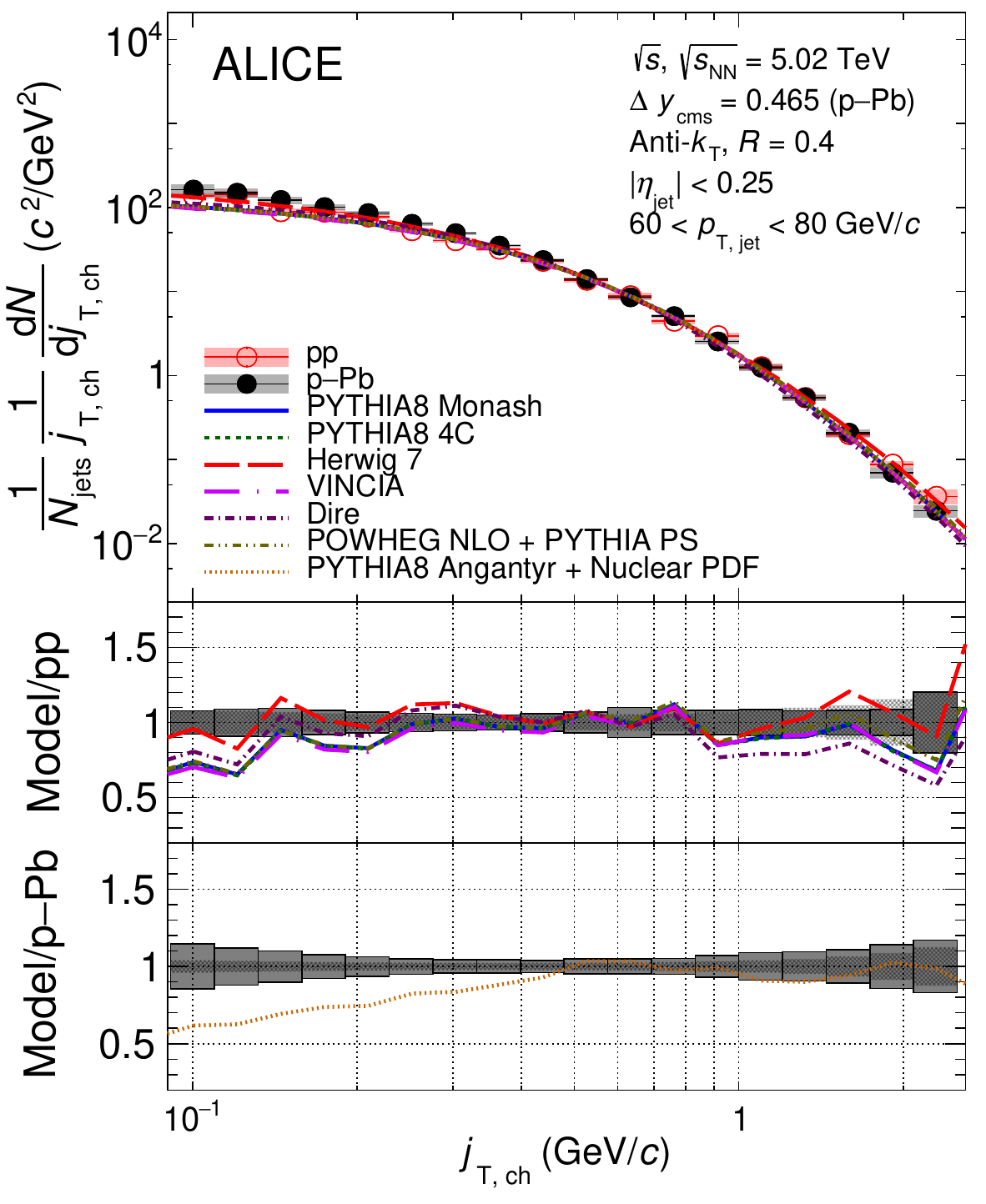}
  \caption{The $\jt$ distribution in $\pPb$ collisions at $\sqrtSnnE{5.02}$ for jets with transverse momentum in \unit[$60<\ptjet<80$]{GeV/$c$}. The measured data are compared to calculations by theoretical models in $\pp$ collisions at \unit[$\sqrt{s}=5.02$]{TeV}.} 
  \label{fig:jtwithmodels}
  \end{center}
\end{figure}

Figure~\ref{fig:jtwithmodels} shows the $\jt$ distribution in $\pPb$ collisions at $\sqrtSnnE{5.02}$ for jets with \unit[$60<\ptjet<80$]{GeV/$c$} compared with expectations from various generators in $\pp$ collisions at \unit[$\sqrt{s}=5.02$]{TeV}. PYTHIA 8 based models (PYTHIA 8.3) and Herwig (Herwig 7.2) handle both the showering process and hadronisation differently. PYTHIA 8 uses the Lund string model~\cite{lundString} to perform the hadronisation stage. Herwig uses a cluster model for the hadronisation~\cite{herwigManual,herwig7releaseNote}. PYTHIA 8 has $\pt{}$-ordered showers by default while Herwig implements a parton shower using the coherent branching algorithm~\cite{Gieseke:2003rz}, which has angular ordering as a central feature. 
The $\pt{}$-ordering in a PYTHIA 8 shower is a compromise~\cite{Sjostrand:2004ef}: ordering in the $\pt{}$ at splitting ensures the ordering in the hardness and also effectively favours large angles. Herwig describes the $\jt$ distribution better than other models for the whole $\jt$ region. Other PYTHIA 8 based models describe the data at high $\jt$ but not in the low $\jt$ region. The results for the other $\ptjet$ intervals are reported in Figs.~B1, B2 and B3 that derive the same conclusion. Models describe the data better as $\ptjet$ increases in pp collisions. This is also true at higher $\jt$, however, models underestimate the data at lower $\jt$ consistently for all $\ptjet$ ranges in p--Pb collisions.

PYTHIA 8 Monash 2013~\cite{Skands:2014pea} adopted LHC data to constrain the initial-state radiation and multi-parton interaction parameters based on the default parameters of PYTHIA 8 tune 4C~\cite{Corke:2010yf}. There is no clear separation of the $\jt$ distributions originating from the different tunes of PYTHIA 8. 
As of version 8.3 PYTHIA 8 implemented two more shower models as part of the code. 
Those are VINCIA and Dire Showers that are based on the $k_\mathrm{T}$ (transverse momentum of a dipole)-ordered picture of QCD splitting~\cite{Fischer:2016vfv,Hoche:2015sya}. The $\jt$ distributions generated by the two shower models were obtained by using the default parameters of PYTHIA 8 tune 4C. In order to study the effect of the NLO calculation accuracy for the parton showering in PYTHIA 8 (POWHEG NLO + PYTHIA PS), the $\jt$ distribution generated with the combined POWHEG~\cite{Oleari:2010nx} and PYTHIA simulation is also compared to the data. The $\jt$ distributions obtained with the POWHEG NLO calculation and Dire Shower display themselves as upper and lower bounds of the PYTHIA 8 based models for the higher $\jt$ region; however, they are within the systematic uncertainty of the data for the higher $\jt$ region. 
PYTHIA 8 Angantyr extends pp simulation of PYTHIA 8 to the case of heavy-ion collisions~\cite{Bierlich:2018xfw}. PYTHIA 8 Angantyr is used to simulate p--Pb collisions with the nuclear parton distribution function (PDF) EPS09LO~\cite{Eskola:2009uj} for the Pb-ion beam. The resulting $\jt$ distribution is almost the same with those by pp simulations with a proton PDF and it does not describe the data for the lower $\jt$ region at all.     

\begin{figure}[t]
  \begin{center}
  \includegraphics[width=0.8\textwidth]{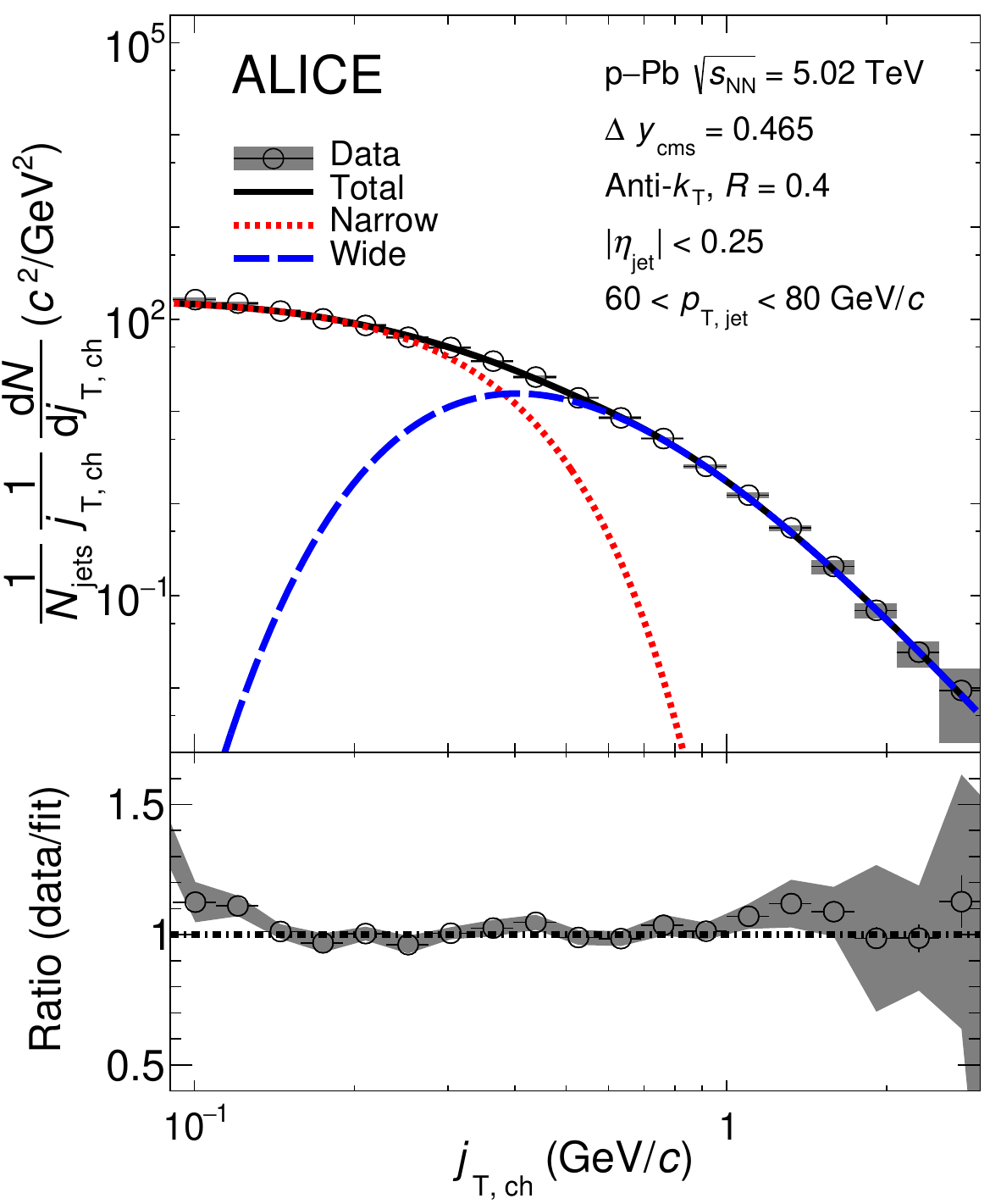}
  \caption{The $\jt$ distribution of charged particles with a two-component fit for \unit[$60<\ptjet<80$]{\GeVc}. The distribution is fitted with the two-component fit described in Sec.~\ref{sec:methods}. }
  \label{fig:fits}
  \end{center}
  \end{figure}

The distributions are fitted with the two-component fit motivated by ~\cite{Acharya:2018edi}. The function forms are given in Eq.~\ref{eq:fit}. An example of the fitted distribution is shown in Fig.~\ref{fig:fits} for \unit[$60<\ptjet<80$]{\GeVc}. The Gaussian term corresponds to the narrow part that can be associated with the hadronisation process, while the inverse gamma corresponds to the wide component characterising the QCD shower. The $\jt$ distributions are described well by the two-component model fit.
The corresponding statistical uncertainties are calculated via the general error propagation formulas in Eq.~\ref{eq:rmserr}

\begin{equation}
\delta \sqrt{\left<\jt^2\right>} = \sqrt{2}\delta B_1 ~~~\textrm{and}~~~ \sqrt{ \left( \frac{(5-2B_4)B_5 \delta B_4}{(2(B_4-2)(B_4-3))^{\frac{3}{2}}} \right)^2 + \left(\frac{\delta B_5}{\sqrt{\left(B_4-2\right)\left(B_4-3\right)}}\right)^2 }
\label{eq:rmserr}
\end{equation}

for the narrow and wide component RMS values, respectively. 

\begin{figure}[t]
  \begin{center}
  \includegraphics[width=0.8\textwidth]{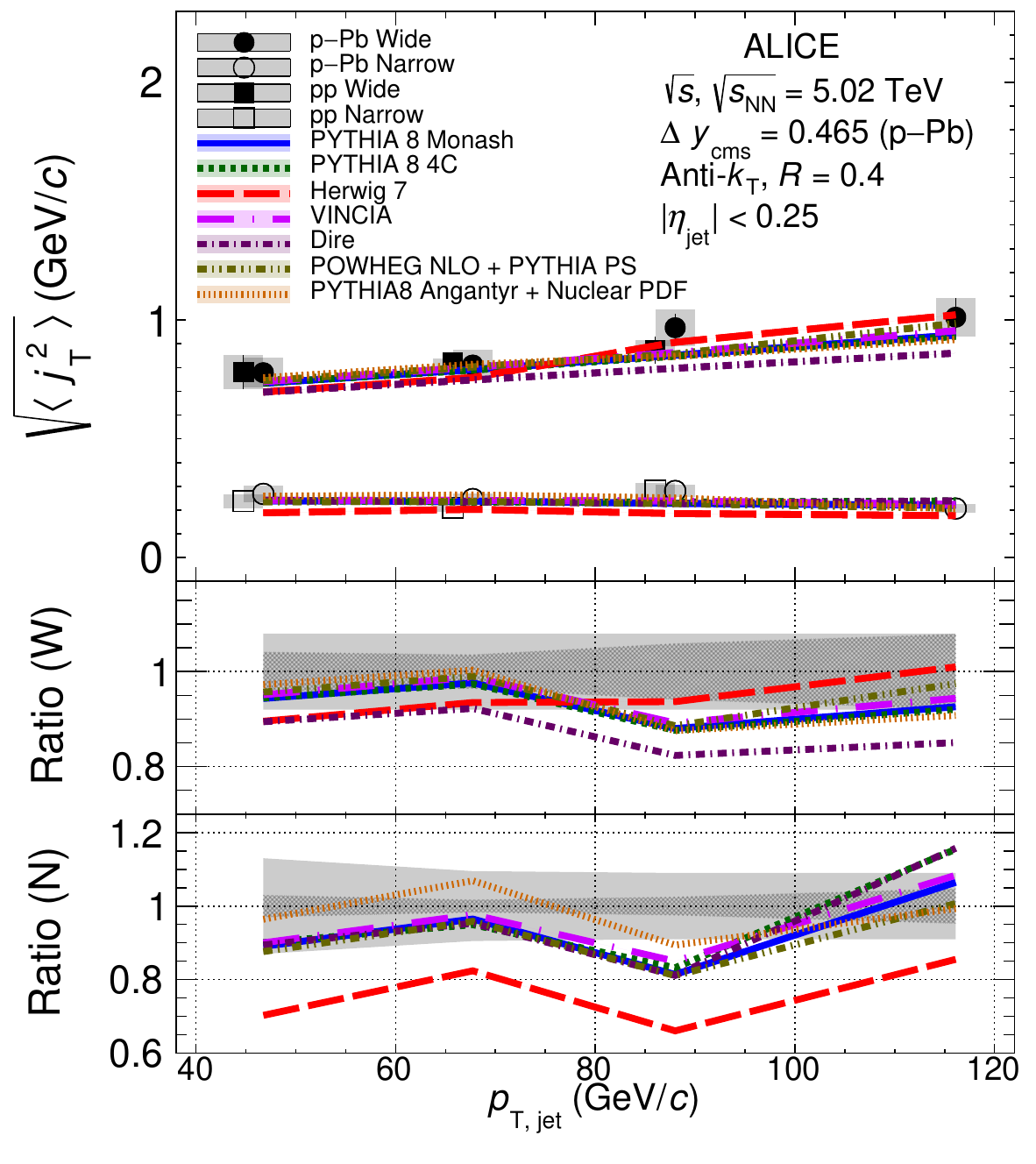}
  \caption{RMS values extracted from the fits for the Gaussian (narrow) and inverse gamma (wide) components. The middle and bottom plots show ratios of models to data for the wide and narrow components, respectively. The grey filled bands with (without) a hatched line in the ratio plots represent the statistical (systematic) uncertainties of the p--Pb data. Note that pp data points are shifted by -2 GeV/$c$ on the horizontal axis to be distinguished from p--Pb data points. }
  \label{fig:rms_models}
  \end{center}
\end{figure}

The widths of the $\jt$ distributions are determined as a function of the transverse momentum of jet. The RMS $\left(\rms{\jt}\,\right)$ values for the two components are shown in Fig.~\ref{fig:rms_models} along with comparisons to Monte Carlo simulations. There is clear separation in the width of the wide and narrow components of the $\jt$ distributions. The RMS values of the wide component are 3-4 times larger than the narrow component RMS. 
The wide component RMS shows an increasing trend with increasing $\ptjet$ that is parameterised by a linear function as $\rms{\jt} = 0.005\,(\pm0.004) \times \ptjet + 0.497\,(\pm0.255)$, while the narrow component RMS stays constant with the fitted value of $0.253\,(\pm0.009)$. Both of these trends are qualitatively consistent with the results in the dihadron $\jt$ analysis~\cite{Acharya:2018edi}. 

All models except for Herwig describe the RMS values relatively well for the narrow RMS component. For the wide RMS component Herwig describes the data best as $\ptjet$ increases. Dire Shower shows clearly lower values compared to data up to 18\% for the wide RMS components. Other PYTHIA 8 based models show a good description for the lower $\jt$ region, however, they underestimate the data for the higher $\ptjet$ region. 

\begin{figure}[htb]
\centering
\includegraphics[width=0.8\textwidth]{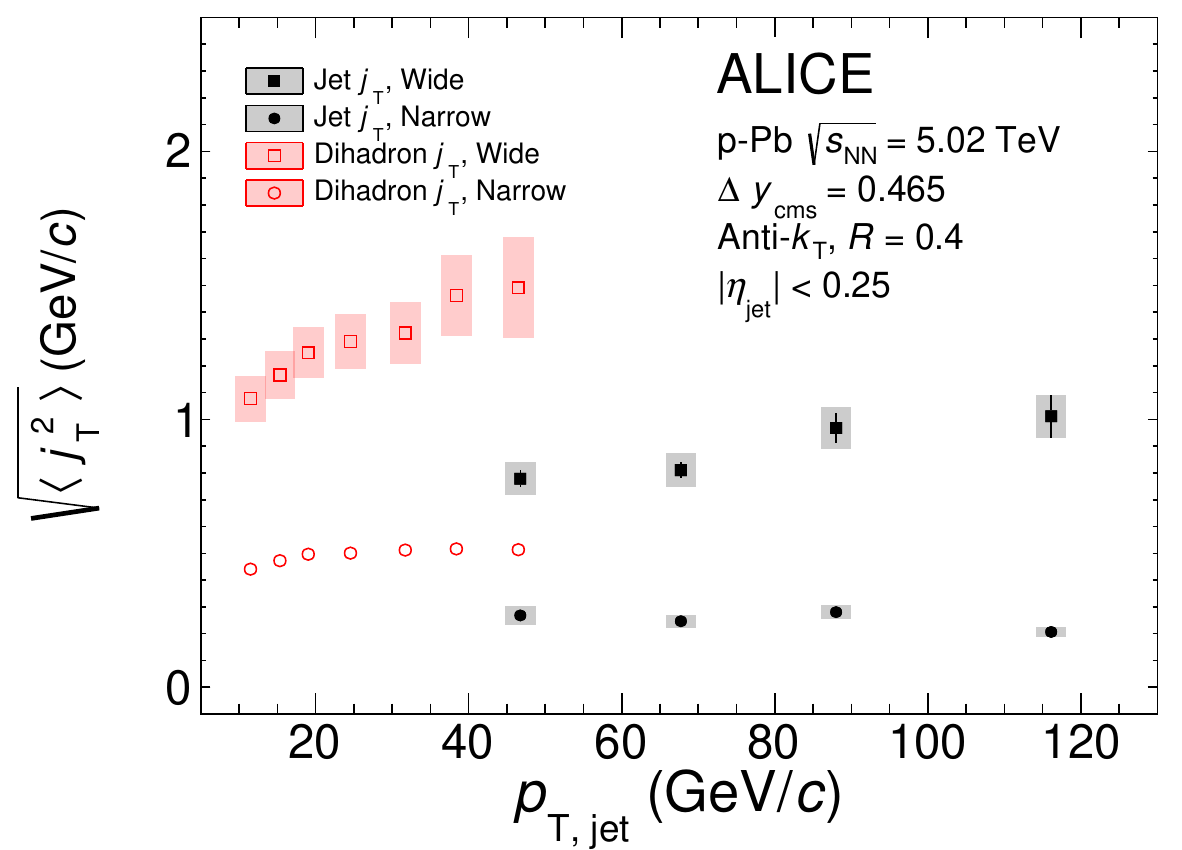}
\caption{Comparison of results from the jet-based and dihadron-based $\jt$ analyses~\cite{Acharya:2018edi}. Ranges of dihadron trigger $\pt{}$ ($\pttrigger$) are converted to corresponding $\ptjet$ ranges using observed mean $\ptjet$ values in $\pttrigger$ bins. Dihadron results are shown for $0.2 < x_{||} < 0.4$, where $x_{||}$ is the longitudinal component fraction of the associated track momentum with respect to the momentum of the trigger track. The difference of the two analyses originates from the different kinematic selections and the choice of the axis used for the $\jt$ calculation. See text for more details. }  
\label{fig:DihadronComparison}
\end{figure}

\section{Discussion}
The comparison with the results from the dihadron analysis~\cite{Acharya:2018edi} performed for the same collision system and energy is shown in Fig.~\ref{fig:DihadronComparison}. Different $p_\mathrm{T}$ regions of leading particles used in the dihadron analysis are converted to the corresponding average momentum of the jets which contain those leading particles. The wide and narrow components of the dihadron results are for $0.2 < x_{||} < 0.4$, where $x_{||}$ is the projection of the momentum of the associated track to that of the trigger particles.  
Wide component RMS values tend to increase with increasing \pttrigger\ and \ptjet, whereas narrow component RMS values of both results show a weak dependence on \ptjet\ above $20\,${~GeV/$c$}. The trends are similar for dihadron and jet $\jt$ results. However, the RMS values of the dihadron analysis are larger than those for the jet analysis both for the narrow and wide components. 

The difference in the narrow and wide RMS components can be explained by the following two factors. The first one is due to the different kinematic selections on the charged particles in the same jet from which the $\jt$ values are calculated. The other one is due to the choice of the axis used for the $\jt$ calculation. 
In the dihadron analysis $\jt$ is calculated for all near-side tracks if the associated tracks satisfy the condition $\vec{p}_\mathrm{leading}\times\vec{p}_\mathrm{a}>0$. Here $\vec{p}_\mathrm{leading}$ and $\vec{p}_\mathrm{a}$ are the momentum vectors of the leading and associated tracks, respectively. 
Thus, the kinematical limit $j_\mathrm{T,max}$ can be larger in the dihadron analysis than in the jet analysis in which only particles in a cone with $R=0.4$ are considered. 

\begin{figure}[t]
  \centering
 % \begin{subfigure}{0.48\textwidth}
  %\includegraphics[width=\linewidth]{figures/Figure6_1}
  \subfigure[]{\includegraphics[width=0.48\textwidth]{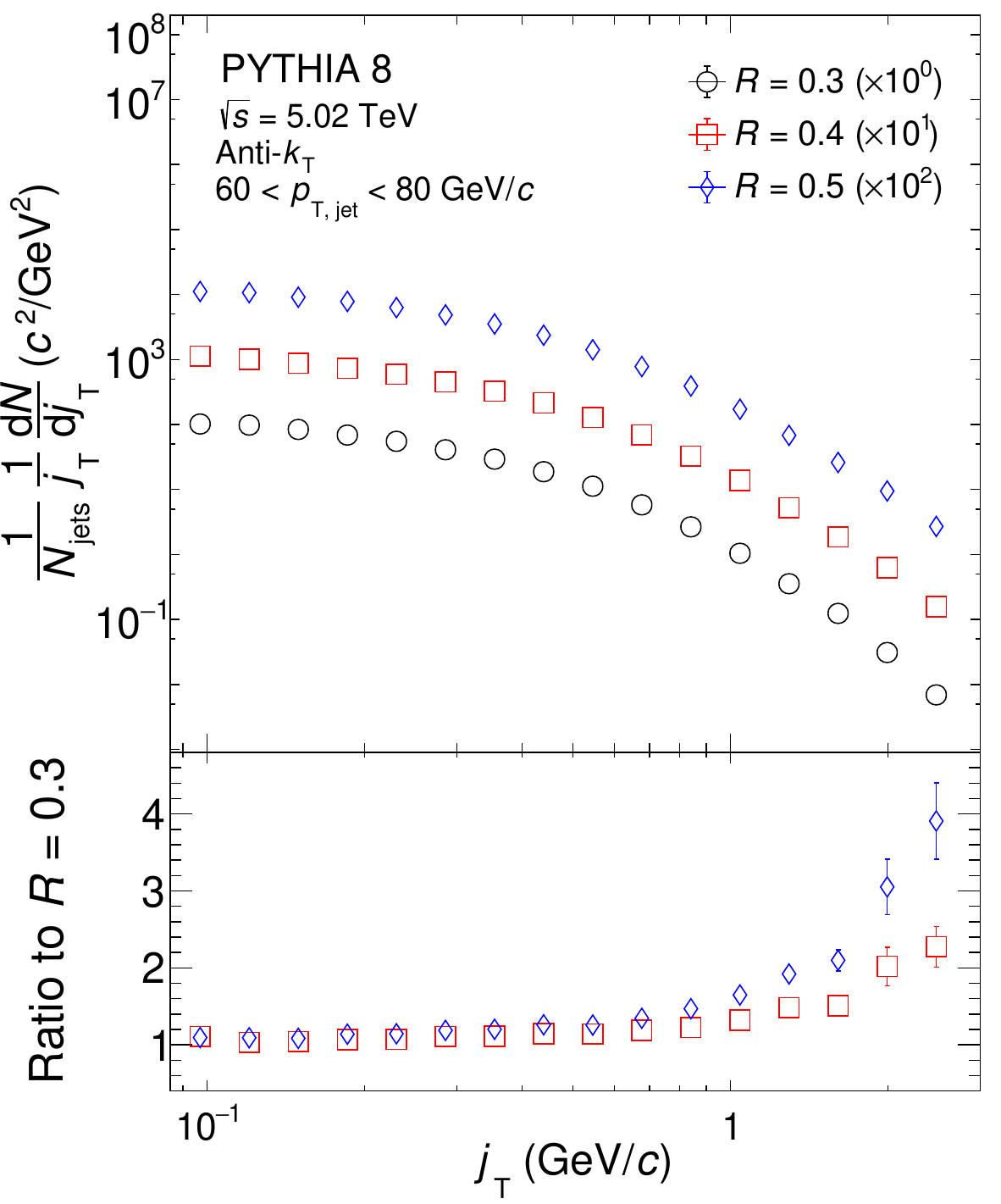}}
  %\end{subfigure}
 % \begin{subfigure}{0.48\textwidth}
  %\includegraphics[width=\linewidth]{figures/Figure6_2}
  %\caption{}
 % \end{subfigure}
\subfigure[]{\includegraphics[width=0.48\textwidth]{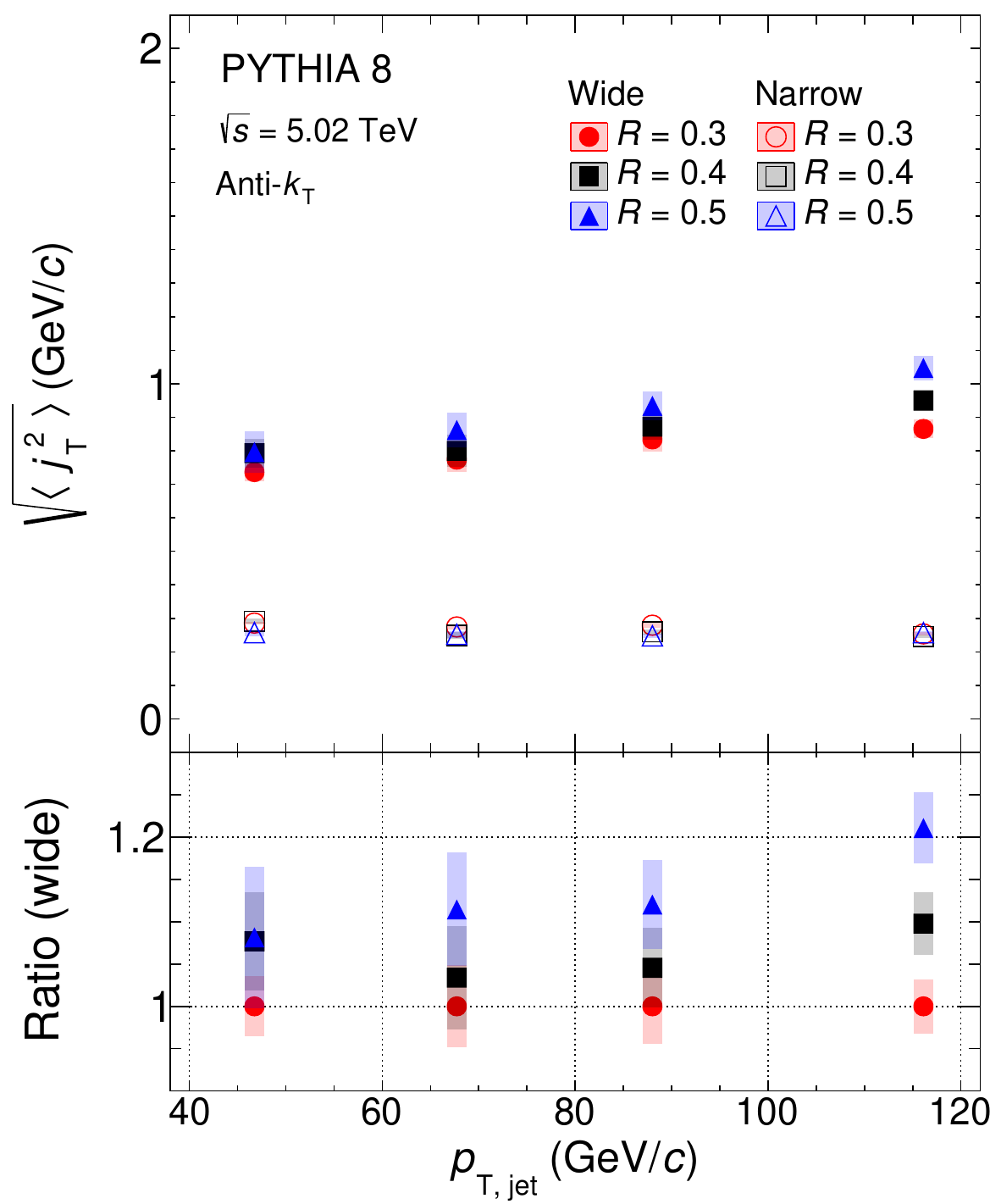}}
  \caption{The effect of changing the $R$ parameter in jet finding on $\jt$ distributions obtained with PYTHIA 8 simulations. Comparison of  (a) $\jt$ signal distributions for different $R$ parameters and their ratios to that of $R=0.3$ and (b) RMS values of the wide and narrow components and their ratios to that of $R=0.3$ for the wide component only.}

  \label{fig:Rcomparison}
\end{figure}

The effect of the $R$ parameter choice and \ptjet\ dependence on $\jt$ was studied using PYTHIA 8 and the results are shown in Fig.~\ref{fig:Rcomparison}a. The usage of a fixed cone sets stringent limits on the possible $\jt$ values. Increasing the cone size loosens these limits and allows for higher $\jt$ values. 
The effect on the wide and narrow components of the $\jt$ distributions for PYTHIA 8 is shown in Fig.~\ref{fig:Rcomparison}b, where the wide component RMS gets larger by about 10\% when going from $R=0.3$ to 0.4 and from 0.4 to 0.5, indicating that the kinematic limit introduced by increasing $R$ results in a widening of the $\jt$ distribution. For the narrow component the effect is relatively small and they appear independent of the $R$ parameter and $\ptjet$. There can also be a broadening effect for jets caused by the increasing gluon jet fraction as the kinematical limit increases~\cite{Larkoski:2014pca}. Additionally, there is an effect originating from the kinematic cut on $x_{||}$ values in the dihadron analysis that can alter the $\jt$ distributions -- but that is not further investigated here.

\begin{figure}[hbt!]
    \centering
    \includegraphics[width=\textwidth]{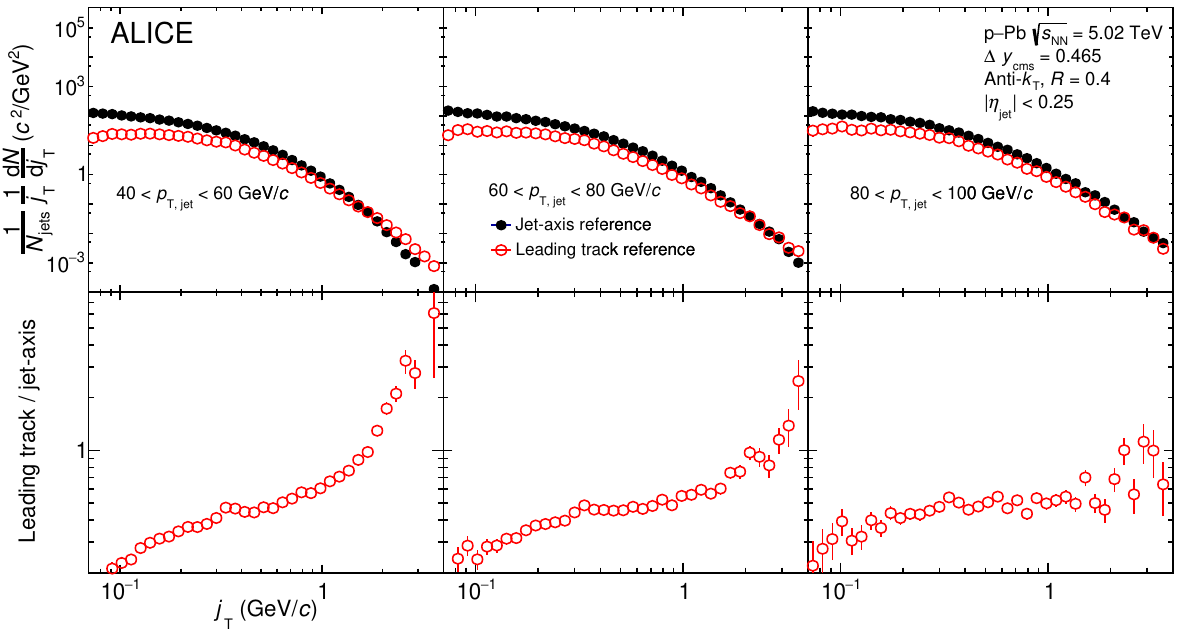}
    \caption{ The $\jt$ distributions with respect to the leading track momentum (leading track reference) and the jet axis (jet-axis reference) within the same jet for three different \ptjet\ intervals with $R=0.4$.}
    \label{fig:RefComparison}
 
\end{figure}

It is worth noting that the leading-track momentum vector provides an imperfect estimate of the jet axis. Because the leading track in general is at an angle compared to the jet axis, the resulting $\jt$ values based on the leading track are biased from the axis of the jet. Practically, the jet axis found by the jet finding algorithm tends to minimise the $\jt$ of jet constituents. Moreover, in the dihadron correlation analysis the usage of the leading hadron as the trigger particle imposes a trigger bias favouring quark jets resulting in jet narrowing.  The impact of the different axes adopted in the two analyses is investigated by measuring $\jt$ with respect to the leading track momentum (leading track reference), instead of the jet axis (jet-axis reference) within the same jet for $R=0.4$. The results are shown in Fig.~\ref{fig:RefComparison}. The widths of the $\jt$ distributions for the jet-axis reference overall are smaller than those of the leading track reference. The bias of the choice of axis becomes small as $\ptjet$ increases.
As shown in the bottom panels, the ratios of the distributions increase monotonically, implying that the leading track reference makes both the wide and narrow components wider as the ratio distributions show a monotonic increase. 

Dihadron $\jt$ distributions~\cite{Acharya:2018edi} are compared to those of jet $\jt$. Although a direct comparison between jet and dihadron $\jt$ measurements is not possible because of the effects of the different kinematic selection and choice of the axis, RMS values of the wide and narrow components can be quantitatively understood by considering the good agreement between PYTHIA and data.

\section{Conclusion}
\label{sec:summary}
In this work the jet fragmentation transverse momentum ($\jt$) distribution of charged particles $\frac{1}{N_{\mathrm{jets}}}\frac{\mathrm{d}N}{\jtch \mathrm{d}\jtch}$ is studied using jet reconstruction in pp and $\pPb$ collisions at $\sqrtS$, $\sqrtSnnE{5.02}$. The $\jt$ distributions of charged particles in p--Pb collisions become wider as the jet transverse momentum \ptjet\ increases. This is understood as an effect of the reduction of the kinematical limit with increasing \ptjet, allowing for higher $\jt$ values. The $\jt$ distribution in p--Pb collision is compared with that in pp collisions for jet transverse momentum in \unit[$40<\ptjet<100$]{GeV/$c$}, which shows no clear modification of the $\jt$ distribution for the p--Pb collision system. No significant cold nuclear matter effects are observed in the previous and current $\jt$ measurements using dihadron correlations~\cite{Acharya:2018edi} and jet reconstruction. For the jet study, higher statistics in pp collisions for both minimum bias and EMCal trigger is demanded to interpret the effect in lower $\jt$ and higher $\ptjet$. The $\jt$ distributions in p--Pb collisions are compared with various parton shower and fragmentation models. All models describe the data well for the higher $\jt$ region, while they underestimate the data by about 20\% and 40\% at lower $\jt$ in pp and p--Pb collisions, respectively. 

Two distinct components of the jet fragmentation transverse momentum $\jt$ are extracted for narrow and wide contributions to quantify the $\jt$ distribution further in pp and p--Pb collisions. The width of the narrow component has only a weak dependence on jet transverse momentum, while that of the wide component increases with increasing jet transverse momentum. The results are qualitatively consistent as a function of $\ptjet$ with the previous $\jt$ study performed with dihadron correlations~\cite{Acharya:2018edi}. We also present a comparison to PYTHIA 8 (PYTHIA 8.3) and Herwig (Herwig 7.2) simulations to figure out if the two distinct components are described well by models or differences are present. 
For the wide component, Herwig and PYTHIA 8 based models slightly underestimate the data for the higher jet transverse momentum region. 
For the narrow component, the measured trends are successfully described by all models except for Herwig. This is opposite to the case of the $\jt$ distributions at lower $\jt$ where the narrow component corresponds. 
This indicates that the shape of the $\jt$ distribution in models is also important to describe the data.

In addition to the result in p--Pb collisions, a high statistics in pp collisions will further constrain predictions in model calculations for jet fragmentation and hadronisation. 
Future studies of the $\jt$ distribution performed differentially in the longitudinal momentum fraction $z$ can be used to constrain transverse-momentum dependent fragmentation functions~\cite{Kang:2017glf}.

\appendix

%%%%% acknowledgements
%\newenvironment{acknowledgement}{\relax}{\relax}
%\begin{acknowledgement}
%\section*{Acknowledgements}
%We wish to thank Torbjörn Sjöstrand for his help in defining a di-gluon initial state in \textsc{Pythia}~8.
%\input{acknowledgements.tex}    %%%%%%% done by webmaster team
%\end{acknowledgement}
\newenvironment{acknowledgement}{\relax}{\relax}
\begin{acknowledgement}
\section*{Acknowledgements}
% Version: 2020-11-02

The ALICE Collaboration would like to thank all its engineers and technicians for their invaluable contributions to the construction of the experiment and the CERN accelerator teams for the outstanding performance of the LHC complex.
The ALICE Collaboration gratefully acknowledges the resources and support provided by all Grid centres and the Worldwide LHC Computing Grid (WLCG) collaboration.
The ALICE Collaboration acknowledges the following funding agencies for their support in building and running the ALICE detector:
A. I. Alikhanyan National Science Laboratory (Yerevan Physics Institute) Foundation (ANSL), State Committee of Science and World Federation of Scientists (WFS), Armenia;
Austrian Academy of Sciences, Austrian Science Fund (FWF): [M 2467-N36] and Nationalstiftung f\"{u}r Forschung, Technologie und Entwicklung, Austria;
Ministry of Communications and High Technologies, National Nuclear Research Center, Azerbaijan;
Conselho Nacional de Desenvolvimento Cient\'{\i}fico e Tecnol\'{o}gico (CNPq), Financiadora de Estudos e Projetos (Finep), Funda\c{c}\~{a}o de Amparo \`{a} Pesquisa do Estado de S\~{a}o Paulo (FAPESP) and Universidade Federal do Rio Grande do Sul (UFRGS), Brazil;
Ministry of Education of China (MOEC) , Ministry of Science \& Technology of China (MSTC) and National Natural Science Foundation of China (NSFC), China;
Ministry of Science and Education and Croatian Science Foundation, Croatia;
Centro de Aplicaciones Tecnol\'{o}gicas y Desarrollo Nuclear (CEADEN), Cubaenerg\'{\i}a, Cuba;
Ministry of Education, Youth and Sports of the Czech Republic, Czech Republic;
The Danish Council for Independent Research | Natural Sciences, the VILLUM FONDEN and Danish National Research Foundation (DNRF), Denmark;
Helsinki Institute of Physics (HIP), Finland;
Commissariat \`{a} l'Energie Atomique (CEA) and Institut National de Physique Nucl\'{e}aire et de Physique des Particules (IN2P3) and Centre National de la Recherche Scientifique (CNRS), France;
Bundesministerium f\"{u}r Bildung und Forschung (BMBF) and GSI Helmholtzzentrum f\"{u}r Schwerionenforschung GmbH, Germany;
General Secretariat for Research and Technology, Ministry of Education, Research and Religions, Greece;
National Research, Development and Innovation Office, Hungary;
Department of Atomic Energy Government of India (DAE), Department of Science and Technology, Government of India (DST), University Grants Commission, Government of India (UGC) and Council of Scientific and Industrial Research (CSIR), India;
Indonesian Institute of Science, Indonesia;
Istituto Nazionale di Fisica Nucleare (INFN), Italy;
Institute for Innovative Science and Technology , Nagasaki Institute of Applied Science (IIST), Japanese Ministry of Education, Culture, Sports, Science and Technology (MEXT) and Japan Society for the Promotion of Science (JSPS) KAKENHI, Japan;
Consejo Nacional de Ciencia (CONACYT) y Tecnolog\'{i}a, through Fondo de Cooperaci\'{o}n Internacional en Ciencia y Tecnolog\'{i}a (FONCICYT) and Direcci\'{o}n General de Asuntos del Personal Academico (DGAPA), Mexico;
Nederlandse Organisatie voor Wetenschappelijk Onderzoek (NWO), Netherlands;
The Research Council of Norway, Norway;
Commission on Science and Technology for Sustainable Development in the South (COMSATS), Pakistan;
Pontificia Universidad Cat\'{o}lica del Per\'{u}, Peru;
Ministry of Science and Higher Education, National Science Centre and WUT ID-UB, Poland;
Korea Institute of Science and Technology Information and National Research Foundation of Korea (NRF), Republic of Korea;
Ministry of Education and Scientific Research, Institute of Atomic Physics and Ministry of Research and Innovation and Institute of Atomic Physics, Romania;
Joint Institute for Nuclear Research (JINR), Ministry of Education and Science of the Russian Federation, National Research Centre Kurchatov Institute, Russian Science Foundation and Russian Foundation for Basic Research, Russia;
Ministry of Education, Science, Research and Sport of the Slovak Republic, Slovakia;
National Research Foundation of South Africa, South Africa;
Swedish Research Council (VR) and Knut \& Alice Wallenberg Foundation (KAW), Sweden;
European Organization for Nuclear Research, Switzerland;
Suranaree University of Technology (SUT), National Science and Technology Development Agency (NSDTA) and Office of the Higher Education Commission under NRU project of Thailand, Thailand;
Turkish Atomic Energy Agency (TAEK), Turkey;
National Academy of  Sciences of Ukraine, Ukraine;
Science and Technology Facilities Council (STFC), United Kingdom;
National Science Foundation of the United States of America (NSF) and United States Department of Energy, Office of Nuclear Physics (DOE NP), United States of America.
\end{acknowledgement}

%%%%%%%% Bibliography (In case of using bibtex generate the bbl requested by arXiv)
\bibliographystyle{utphys}   % Remember we use title in the biblio
\bibliography{biblio}

\newpage

\appendix

\section{The ALICE Collaboration}
\label{app:collab}
\begin{flushleft}

\bigskip 

S.~Acharya$^{\rm 142}$, 
D.~Adamov\'{a}$^{\rm 97}$, 
A.~Adler$^{\rm 75}$, 
J.~Adolfsson$^{\rm 82}$, 
G.~Aglieri Rinella$^{\rm 35}$, 
M.~Agnello$^{\rm 31}$, 
N.~Agrawal$^{\rm 55}$, 
Z.~Ahammed$^{\rm 142}$, 
S.~Ahmad$^{\rm 16}$, 
S.U.~Ahn$^{\rm 77}$
% ...

\bigskip

\bigskip 

\textbf{\Large Affiliation Notes}

\bigskip 

% ...
$^{\rm V}$ Also at: Institute of Theoretical Physics, University of Wroclaw, Poland\\

\bigskip

\bigskip 

\textbf{\Large Collaboration Institutes}

\bigskip 

% ...
$^{16}$ Department of Physics, Aligarh Muslim University, Aligarh, India\\
% ...
$^{31}$ Dipartimento DISAT del Politecnico and Sezione INFN, Turin, Italy\\
% ...
$^{35}$ European Organization for Nuclear Research (CERN), Geneva, Switzerland\\
% ...
$^{55}$ INFN, Sezione di Bologna, Bologna, Italy\\
% ...
$^{75}$ Johann-Wolfgang-Goethe Universit\"{a}t Frankfurt Institut f\"{u}r Informatik, Fachbereich Informatik und Mathematik, Frankfurt, Germany\\
% ...
$^{77}$ Korea Institute of Science and Technology Information, Daejeon, Republic of Korea\\
% ...
$^{82}$ Lund University Department of Physics, Division of Particle Physics, Lund, Sweden\\
% ...
$^{97}$ Nuclear Physics Institute of the Czech Academy of Sciences, \v{R}e\v{z} u Prahy, Czech Republic\\
% ...
$^{142}$ Variable Energy Cyclotron Centre, Homi Bhabha National Institute, Kolkata, India\\
% ...

\bigskip 

\end{flushleft} 
 
\clearpage
\section{ Comparison of the $j_\mathrm{T}$ distributions with models for other $\ptjet$ regions }

\begin{figure}[!htb]
  \begin{center}
  \includegraphics[width=0.45\textwidth]{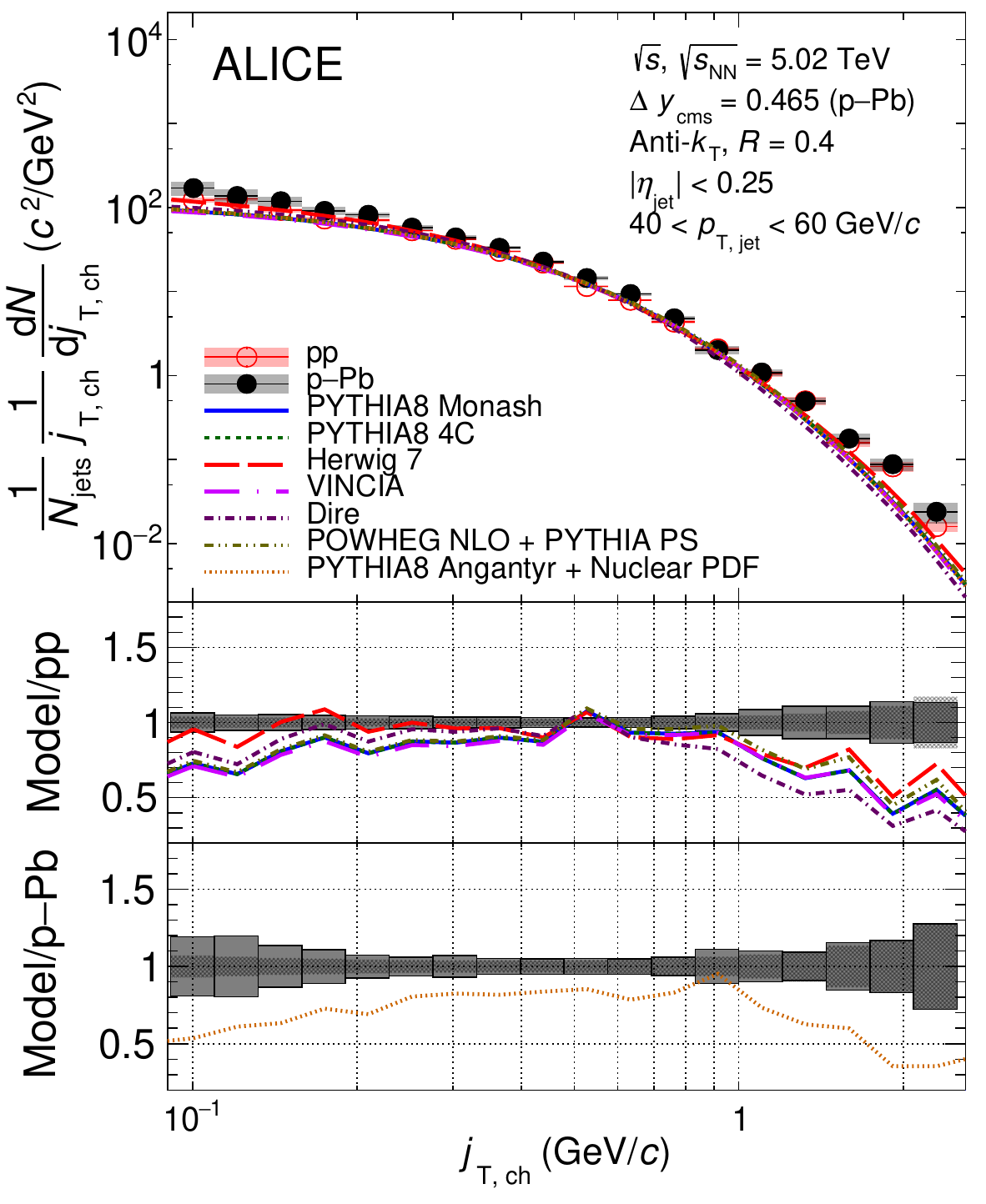}
  \caption{The $\jt{}$ distribution in pp and $\pPb$ collisions at $\sqrt{s},\,\sqrtSnnE{5.02}$ for \unit[$40<\ptjet<60$]{GeV/$c$} comparing to theoretical models in pp and p--Pb collisions.}
  %All models are below the data points and located around the lower bound of the error of data.}
  \label{fig:jtwithmodels_4060}
  %\includegraphics[width=0.48\textwidth]{figures/results/JtSignalFinalFitJetPt7.pdf}
  %The Gaussian is responsible for the narrow part coming from the hadronisation process, on the other hand, the inverse gamma is for the wide one originating from the showering process.}
  %Fig2_DrawFinalFits.py in https://github.com/TWSman/JtAnalysis
  \end{center}
  \end{figure}

  \begin{figure}[!htb]
  \begin{center}
  \includegraphics[width=0.45\textwidth]{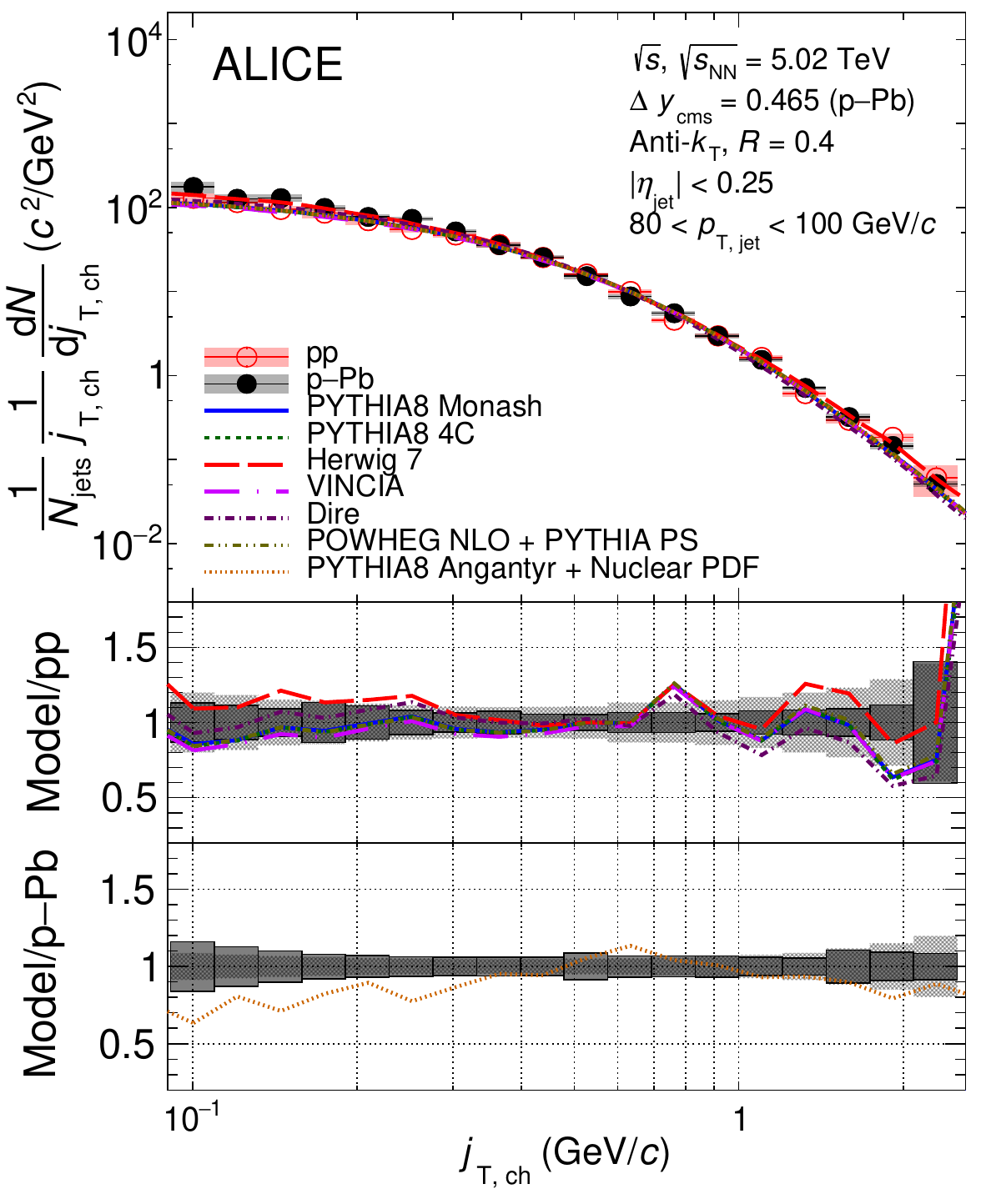}
  \caption{The $\jt{}$ distribution in $\pPb$ collisions at $\sqrt{s},\,\sqrtSnnE{5.02}$ for \unit[$80<\ptjet<100$]{GeV/$c$} comparing to theoretical models in pp and p--Pb collisions. }
  %Almost models describe the data relatively well for \unit[$\jt{}>0.7$]{GeV/$c$} than for \unit[$\jt{}<0.7$]{GeV/$c$}}
  \label{fig:jtwithmodels_80100}
  %\includegraphics[width=0.48\textwidth]{figures/results/JtSignalFinalFitJetPt7.pdf}
  %The Gaussian is responsible for the narrow part coming from the hadronisation process, on the other hand, the inverse gamma is for the wide one originating from the showering process.}
  %Fig2_DrawFinalFits.py in https://github.com/TWSman/JtAnalysis
  \end{center}
  \end{figure}

  \begin{figure}[!htb]
  \begin{center}
  \includegraphics[width=0.5\textwidth]{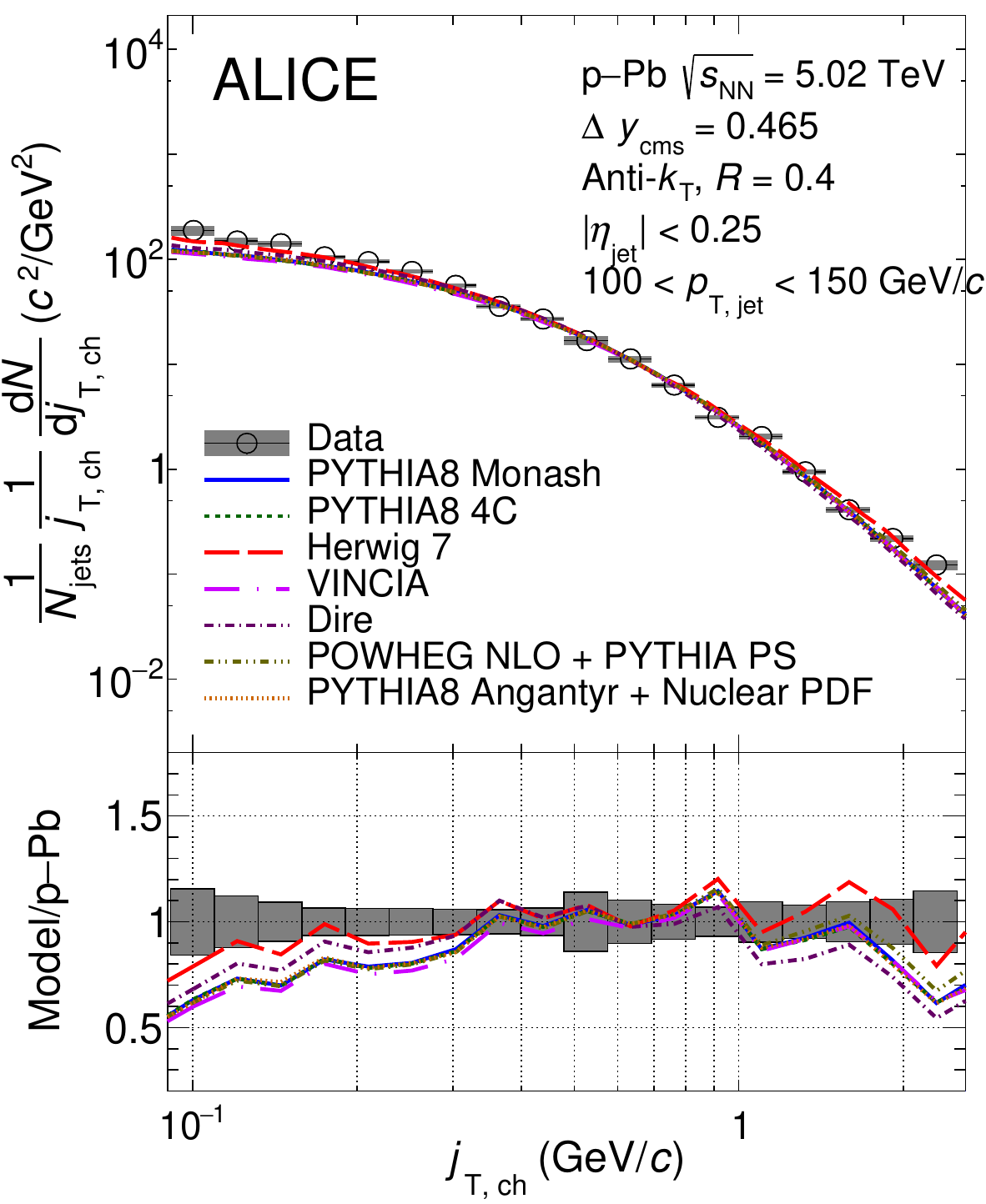}
  \caption{The $\jt{}$ distribution in $\pPb$ collisions at $\sqrtSnnE{5.02}$ for \unit[$100<\ptjet<150$]{GeV/$c$} comparing to theoretical models in pp and p--Pb collisions. }
  %All models describe the data well for the higher $\jt{}$ than the lower $\jt{}$ region. }
  \label{fig:jtwithmodels_4060}
  %\includegraphics[width=0.48\textwidth]{figures/results/JtSignalFinalFitJetPt7.pdf}
  %The Gaussian is responsible for the narrow part coming from the hadronisation process, on the other hand, the inverse gamma is for the wide one originating from the showering process.}
  %Fig2_DrawFinalFits.py in https://github.com/TWSman/JtAnalysis
  \end{center}
  \end{figure}
  \clearpage
  \iffalse
  \section{ Two-component model fit in pp collisions}
    \begin{figure}[!htb]
  \begin{center}
  \includegraphics[width=0.5\textwidth]{figures/Fitting_pp.pdf}
  \caption{The $\jt{}$ distribution of charged particles with a two-component fit for \unit[$40<\ptjet<60$]{\GeVc} in pp collisions. The distribution is fitted with the two-component fit described in Sec.~\ref{sec:methods}. }
    \end{center}
  \end{figure}
  \fi

\end{document}